\documentclass[twocolumn,aps,prb,final,superscriptaddress,floatfix]{revtex4}

\usepackage{amssymb}

\usepackage[pdftex]{graphicx}

\newcommand{\invA}{\mathrm{\AA}^{-1}}

\newcommand{\Gbar}{\overline{\Gamma}}
\newcommand{\Bbar}{\overline{\mathrm{B}}}
\newcommand{\Ybar}{\overline{\mathrm{Y}}}
\newcommand{\Dbar}{\overline{\mathrm{D}}}

\begin{document}

\title{Splitting in the Fermi surface of ZrTe$_{3}$: a surface charge density wave system}

\author{Moritz Hoesch}
\affiliation{European Synchrotron Radiation Facility, 6 rue Jules Horowitz, 38043 Grenoble Cedex, France}

\author{Xiaoyu Cui}
\altaffiliation{{\em Now at} Swiss Light Source, Paul Scherrer Institut, 5232 Villigen PSI, Switzerland}

\author{Kenya Shimada}
\affiliation{Hiroshima Synchrotron Radiation Centre, Hiroshima University, Kagamiyama 2-313, Higashi-Hiroshima 739-0046, Japan}

\author{Corsin Battaglia}
\affiliation{Institut de Physique, Universit\'{e} de Neuch\^{a}tel, CH-2000 Neuch\^{a}tel, Switzerland}

\author{Shin-ichi Fujimori}
\affiliation{Synchrotron Radiation Research Unit, Japan Atomic Energy Agency, Mikazuki, Hyogo 679-5148, Japan}

\author{Helmuth Berger}
\affiliation{Ecole Polytechnique F\'{e}derale de Lausanne, Institut de physique de la mati\`{e}re complexe, 1015 Lausanne, Switzerland}

\date{\today}

\begin{abstract}
The electronic band structure and Fermi surface of ZrTe$_3$ was precisely determined by linearly polarized angle-resolved photoelectron spectroscopy. Several bands and a large part of the Fermi surface are found to be split by 100-200 meV into two parallel dispersions. Band structure calculations reveal that the splitting is due to a change of crystal structure near the surface. The agreement between calculation and experiment is enhanced by including the spin-orbit potential in the calculations, but the spin-orbit energy does not lead to a splitting of the bands. The dispersion of the highly nested small electron pocket that gives rise to the charge density wave is traceable even in the low-temperature gapped state, thus implying that the finite correlation length of the long-wavelength modulation leads to a smearing of the band back-folding.
\end{abstract}

\pacs{}

\maketitle

\section{Introduction}

In a one-dimensional metal the atoms form chains and specific electronic states are confined to travel predominantly along these chains. This leads to a Fermi surface (FS) topography that is characterized by parallel sheets as the band dispersion is significant only along the chain direction. The corresponding FS nesting makes the system susceptible to instabilities such as spin or charge density waves (SDW and CDW), if assisted by a coupling to the respective magnetic or lattice degrees of freedom. Chain-like arrangements are found in a number of crystal structures and CDW phenomena are observed in varied systems such as blue bronze, KCP, (TaSe$_4$)$_2$I and the prismatic chains of NbSe$_3$ and TaS$_3$.\cite{grunerbook} The ideal CDW should be an isolated Peierls chain, but in a crystal a certain coupling between parallel chains is always present. Reduced coupling of adjacent chains is achieved in surface CDW systems such as In/Si(111)\cite{yeom99} and Au/Ge(001).\cite{schaefer08}

To date it is still very difficult to predict whether a material will exhibit a charge density wave. Fermi surface nesting is certainly an important ingredient and can be studied by ground state electronic structure calculations of the unmodulated structure. The formation of a statically modulated CDW ground state requires a certain coupling between the chains so that below the Peierls transition a three-dimensionally ordered ground state can exist. Phonons eventually destroy the CDW order at the Peierls transition temperatures, but on the other hand the electron-phonon coupling EPC is also needed to assist the formation of a static distortion where the lattice degrees of freedom follow the electronic modulation. It is thus highly interesting to study how the CDW reacts to slight changes in the chain structure and coupling, which can be induced by pressure,\cite{yomo05} strain, or, in the case discussed in this paper, by the presence of the surface.

\begin{figure}[b!]
  \centerline{\includegraphics[width = 0.5\textwidth]{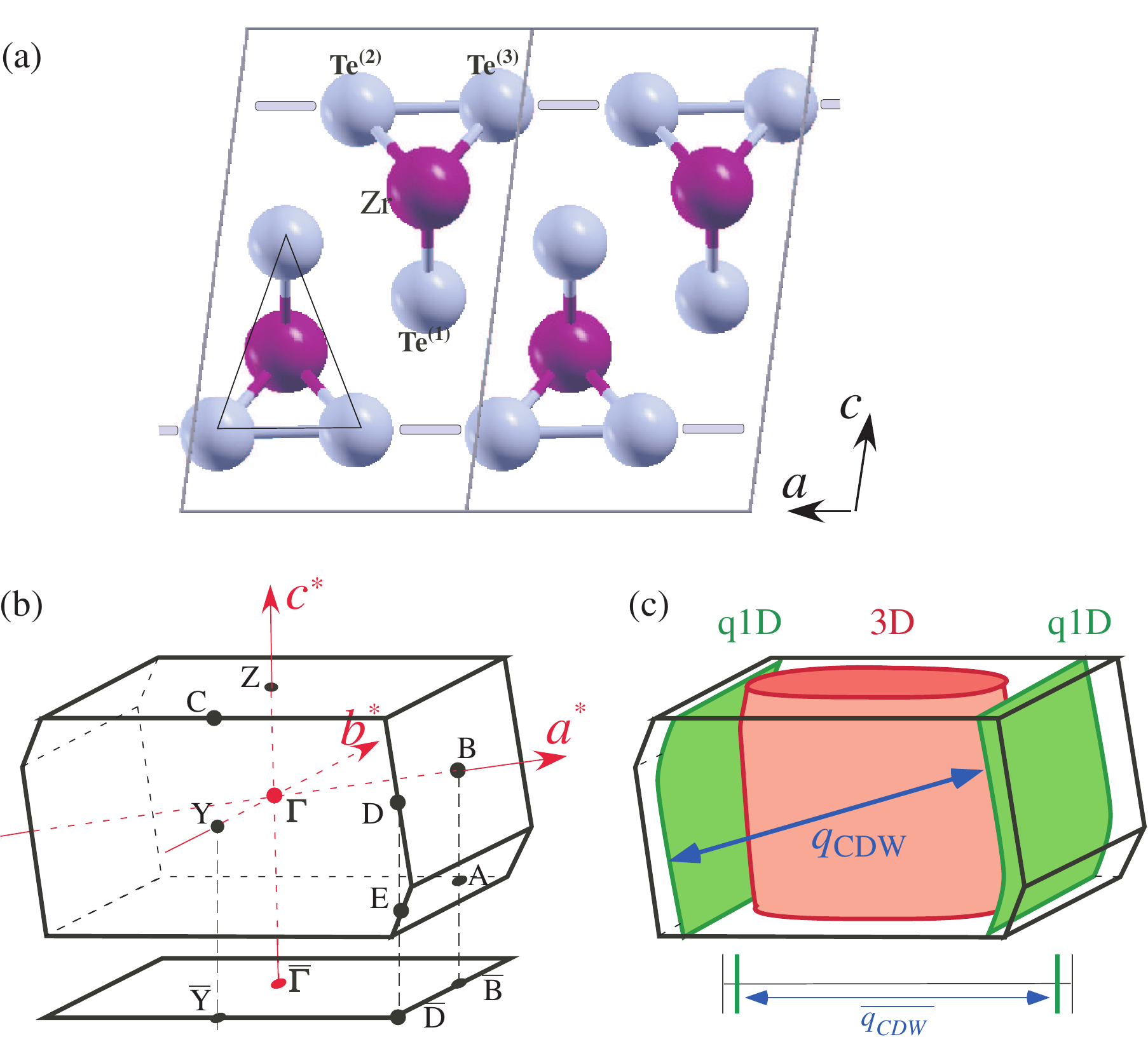}}
  \caption{(a) Projection of the unit cell onto the $a$-$c$ plane of the structure showing the chains in the Te layer adjacent  
   to the van der Waals gap. (b) First Brillouin zone of the monoclinic structure and surface Brillouin zone (SBZ). 
   (c) Schematic representation of the two main sheets of the Fermi surface with the nesting vector. 
   $\overline{q_{CDW}}$ represents the projection of $q_{CDW}$ onto the SBZ.
   }
  \label{struct}
\end{figure}

We present an extended data set of photoelectron spectroscopy data of the layered material ZrTe$_3$. A cleaved surface of a van der Waals bonded layered system is usually assumed to have a bulk-like termination and the absence of dangling bonds makes the formation of surface states unlikely. Thus the surface-sensitive photoelectron signal is often regarded as representative of the bulk electronic structure. Indeed, in ZrTe$_3$ a previous photoelectron spectroscopy study\cite{yokoya05} found the momentum structure of the charge density wave and its temperature dependence in good agreement with the complementary and less surface sensitive optical conductivity data.\cite{perucchi04} We find, however, that our data are not compatible with the expected features of the bulk electronic structure and conclude that the signal, as well as the results of other studies,\cite{yokoya05,starkovicz07,pacile07} represent a special surface system that is modified from the bulk structure, with the interesting consequence that this modified system shows potential differences to the related bulk of ZrTe$_3$.



In the bulk, the CDW transition of ZrTe$_3$ is well characterized. It was first observed as a resistivity anomaly at $T_{CDW}=63$~K for measurements along the crystal $a$ and $c$-directions.\cite{takahashi83} The periodic lattice distortion was observed by diffraction\cite{eaglesham84} and by a giant-Kohn-anomaly behaviour of a soft phonon.\cite{hoesch09} Below $T_{CDW}$ the system is left in a metallic state, where superconducting filaments emerge at very low $T$ ($T_c=2$~K).\cite{takahashi83,yomo05} The opening of a gap in the electronic structure does not affect the entire Fermi surface. In fact, the Fermi surface consists of at least two principal sheets (Fig.~\ref{struct}(c)) as seen in calculations\cite{felser98,stowe98} and experiments.\cite{yokoya05, starkovicz07,kubo07} A central 3D Fermi surface is derived from holes that propagate mostly along the prismatic (ZrTe$_3)_\infty$ chains along the $b$-direction. The elongated cross section in the $a^*$-$b^*$ plane allows for a certain nesting along $b^*$. A much better nesting condition is found in the small electron pocket of the Te 5$p_x$ band which propagates along the Te-Te chains formed by the lateral bonding of adjacent prisms. This quasi one-dimensional (q1D) Fermi surface with principal propagation direction along $a$ was identified to match the observed modulation vector $q_{CDW} = (0.07,\ 0, \ 0.333).$\cite{eaglesham84} Strictly speaking, the q1D FS consists of two sheets due to the inequivalence of the two Te atoms in the Te-Te chains. In the out-of-plane direction $c$ the structure is much more weakly bonded since only van der Waals forces act between layers. The band dispersion is consequently much weaker. Some dispersion has, however, been predicted by all calculations. In particular the q1D Fermi surface can be seen to follow the inclination of the Brillouin zone boundary (BZB) given by the monoclinic $P2_1/m$ structure.\cite{felser98}

 The Te layer adjacent to the van der Waals gap is of special interest. It is formed by the inequivalent but almost equidistant Te$^{(2)}$ and Te$^{(3)}$ atoms and its band structure has been discussed in Ref.~\onlinecite{felser98}. Due to the weak bonding it is the most likely surface termination layer in a cleaved sample. The nesting $\overline{q_{CDW}}$ of the artificial two-dimensional single layer would be along the $\Gbar$-$\Bbar$ direction. 



 
 This paper is organized as follows: in the next section we detail the experimental procedure and computational details. Sect.~\ref{sect_band} shows the measured data and discusses the observed band dispersions and splittings. Sect.~\ref{sect_split} discusses the implications of the observed splitting through comparison with calculations. Sect.~\ref{sect_CDW} reports on the observations concerning the CDW gap. The conclusions are summarized in the final section.
 
\section{Experiment and calculations}

Single crystals of ZrTe$_3$ were grown by vapour-phase transport in the form of large shiny platelets. Photoemission data were acquired from samples freshly cleaved at room temperature by peeling off the top layers with Scotch$^{\mathrm{TM}}$-tape. In an ultrahigh vacuum of typical residual pressure $2\cdot 10^{-10}$~mbar the samples remained clean and sharp spectra could be acquired over a two days period following cleavage. The sample normal and high symmetry directions were determined from the shape of the sample, the direction of the fibres along the $b$ axis and the symmetry of photoemission angular scans. The Fermi level and the total energy resolution ($\Delta E = 17$~meV) were determined from the metallic Fermi cut-off of a polycrystalline gold film.

The angle-resolved photoelectron spectroscopy (ARPES) experiments were performed at the linearly-polarized undulator beamline BL-1 at the Hiroshima Synchrotron Radiation Centre.\cite{shimada01} The experimental station is equipped with a low-temperature two-axis goniometer. The angle-dispersive detection direction of the electron spectrometer is in the plane of linear polarization of the photons. When a mirror plane of the sample is aligned with this plane, the spectra will only contain contributions from orbitals that are even with respect to the mirror plane (selection rule). When the sample is tilted about a horizontal rotation axis, both even and odd states may be detected depending on the magnitude of the tilt. Fermi surface maps have been acquired by sequentially tilting the sample across and away from this symmetric situation to cover the irreducible portions of the two-dimensional in-plane Brillouin zone. Additional data were acquired using a similar set-up with even higher angular resolution at the Swiss Light Source using elliptically polarized light (Figs.~\ref{FSM} and \ref{CDW_gap}).

Mapping of the photoelectron angular distribution to momentum $k_{||}$ was performed according to standard parallel projections. The locus of the Brillouin zone boundaries was determined from the symmetry of the data sets. While ARPES could in principle detect the full three-dimensional momentum vector,\cite{rossnagel01,strocov06} we find that the observed band dispersions agree perfectly for data sets acquired at different photon energies $h\nu =36.7 - 58.8$~eV. Variations are seen only in the relative intensity of features, but not in their binding energies at corresponding $k_{||}$. Our data therefore display a two-dimensional band structure and we denote the high-symmetry points in the surface Brillouin zone as $\Gbar$, $\Ybar$, $\Bbar$ and $\Dbar$ (cf Fig.~\ref{struct}(b)).

Two distinct sample mountings were employed that lead to different orientations of the polarization vector with respect to the sample. With the $b$-axis mounted horizontally the central 3D Fermi surface was accurately mapped since the tilt corresponds to the dispersion along $\Gbar$-$\Bbar$. No mirror plane of bulk ZrTe$_3$ is present  in these scans, but states that are odd in $a$ are suppressed in this sample orientation. With the $b$-axis mounted vertically, the q1D Fermi surface and the $\Bbar$-point (where the two principal sheets approach) are accurately mapped. Here the selection rule implies that at the $\Bbar$ point only even states with respect to the $a$-$c$ plane are observable.

Peak positions, intensities and widths were extracted from the measured data wherever stable peak fitting could be performed using model functions consisting of two or four Lorentzians. In the case of momentum distribution curves (MDC) at constant binding energy, the widths of the Lorentzians were restricted to be equal for equivalent bands and a constant background was included in the fit function. Typical widths were of the order of $0.07\ \invA$ ($0.03\ \invA$ for the higher resolution data sets). In the case of energy distribution curves (EDC) at constant position in momentum space, the resulting sum of Lorentzians and a linear background was multiplied by a Fermi distribution function with a pre-determined Fermi energy and an effective temperature (width of the Fermi cut off) that takes into account the finite energy resolution. Typical widths of EDC features ranged from 70~meV up to 120~meV for very steep bands, where the momentum resolution has an effect on the spectral widths.

{\em Ab initio} band structure calculations were performed using the Wien2k code.\cite{wien2k} For bulk ZrTe$_3$ the crystal structure from the literature\cite{stowe98} was optimized until the forces on all atoms converged below a limit of 2 mRy/Bohr. The augmented plane wave basis was expanded to a wave-number limit $k_{max}$ given by $r\cdot k_{max}=7$, where $r=2.5$~Bohr is the radius of the atomic spheres. The total energy was integrated over a momentum space mesh of $13\times 19\times 7$ points which was reduced according to the crystal symmetries. For a calculation in a slab geometry the unit cell was doubled in the $c$-direction and a vacuum layer of 10 Bohr thickness was inserted. The two-dimensional electronic structure was calculated on a $26\times 40\times 1$ mesh and the atomic positions were relaxed to less than 2 mRy/Bohr resulting in slightly expanded layers. Other calculation parameters were identical to the bulk calculation. Fermi surface cuts were interpolated from the Fermi level crossings of all bands in a regular $40\times 20$ grid in the symmetry-reduced part of the first Brillouin zone.
 

\section{Band dispersions and splittings}
\label{sect_band}

\begin{figure}
\centerline{\includegraphics[width = 0.5\textwidth]{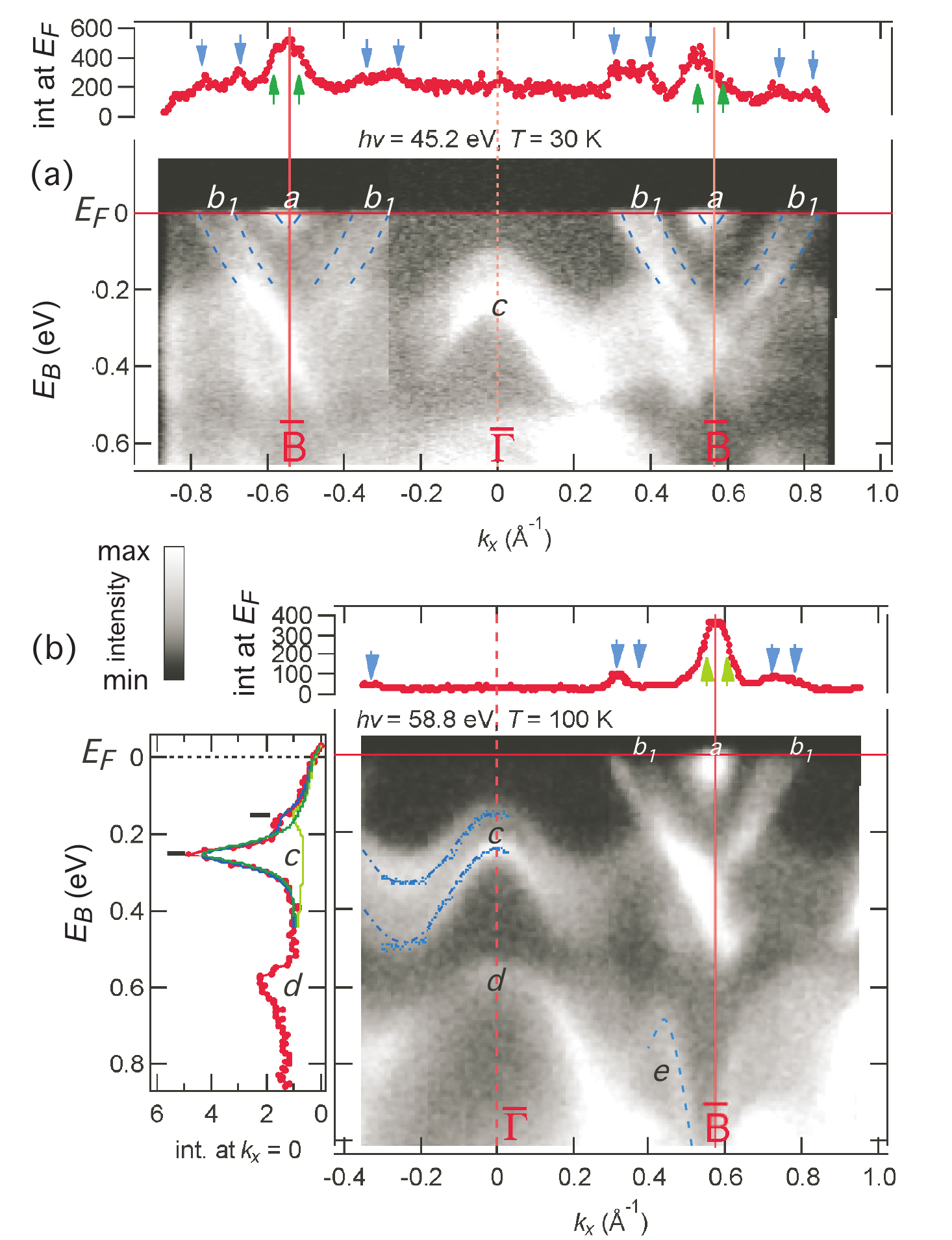}}
\caption{ARPES dispersion maps along the direction $\Gbar-\Bbar$ measured with photon energies (a) $h\nu=42.5$~eV  and (b) $h\nu=58.8$~eV at low temperatures as indicated. Dots in the left section of (b) are peak positions from peak fitting. Dashed lines mark the observed dispersions in selected regions as a guide for the eye. The top of each panel shows the intensity distribution at $E_F$ with the peak positions marked by arrows.}
\label{disp_GB}
\end{figure}

\begin{figure}
\centerline{\includegraphics[width = 0.5\textwidth]{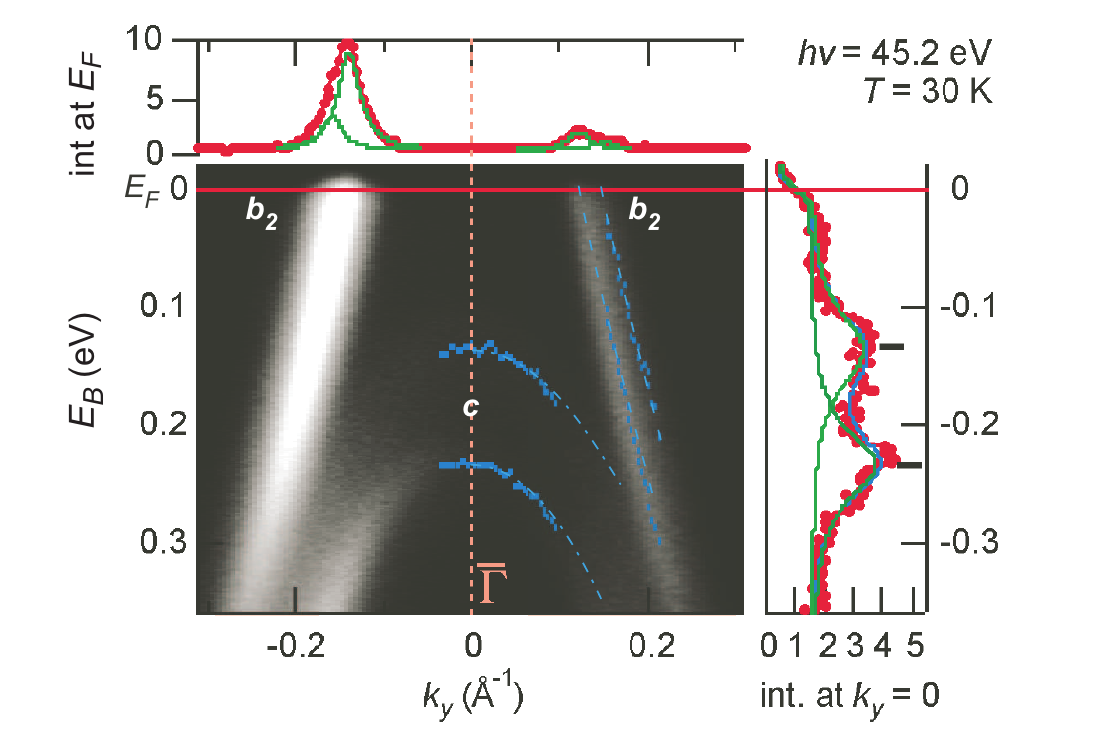}}
\caption{ARPES dispersion map along the direction $\Gbar-\Ybar$ measured with $h\nu=42.5$~eV and $T=30$~K. The top section shows the intensity distribution at $E_F$. Dots are peak positions from fitting as shown in the right section for an EDC at $\Gbar$.} 
\label{disp_GY}
\end{figure}

Dispersion maps at two photon energies across $\Gbar$ in the $a$-$c$ mirror plane are shown in Fig.~\ref{disp_GB}. The photoemission intensity is plotted on a linear grey scale versus the binding energy $E_B$ and the wave vector $k_x$ along $\Gbar$-$\Bbar$. The two data sets differ in sample temperature $T$ and photon energy $h\nu$. Further data sets have been acquired in the photon energy range $h\nu = 36.7 - 56.8$~eV. As the total momentum of the photoelectrons is varied with their energy a signature of the dispersion along $k^{(c^*)}$ is expected. However, except for variations in the relative intensity, the observed bands, their binding energies and dispersion match between all the data sets. Thus the data represent the two-dimensional electronic structure of a ZrTe$_3$ layer.

The top section of each panel shows a cut along $k_x$ (MDC) at $E_F$. The strongest feature is a small pocket {\em "a"} at the $\Bbar$ point just below $E_F$. A second Fermi surface feature {\em "$b_1$"} shows two Fermi crossings at  $k_x = 0.32\ \invA$ and $0.4\ \invA$, together with its symmetry-related copies. It consists of two parallel dispersing bands with a splitting of  $\approx 100$~meV. The third feature {\em "c"} is a completely occupied band with a band apex at $\Gbar$. Two copies with parallel dispersions may again be distinguished, although the relative intensities differ strongly. The left section of Fig.~\ref{disp_GB}(b) shows an EDC at  $\Gbar$, where the two peaks due to this band are highlighted by the resulting Lorentzian functions from peak fitting. The peak positions from this fitting procedure are marked as dots in the panel. The splitting is $\approx 100$~meV at $\Gbar$ and increases to almost 200~meV at the lower band apex.

\begin{figure}[thb!]
\centerline{\includegraphics[width = 0.5\textwidth]{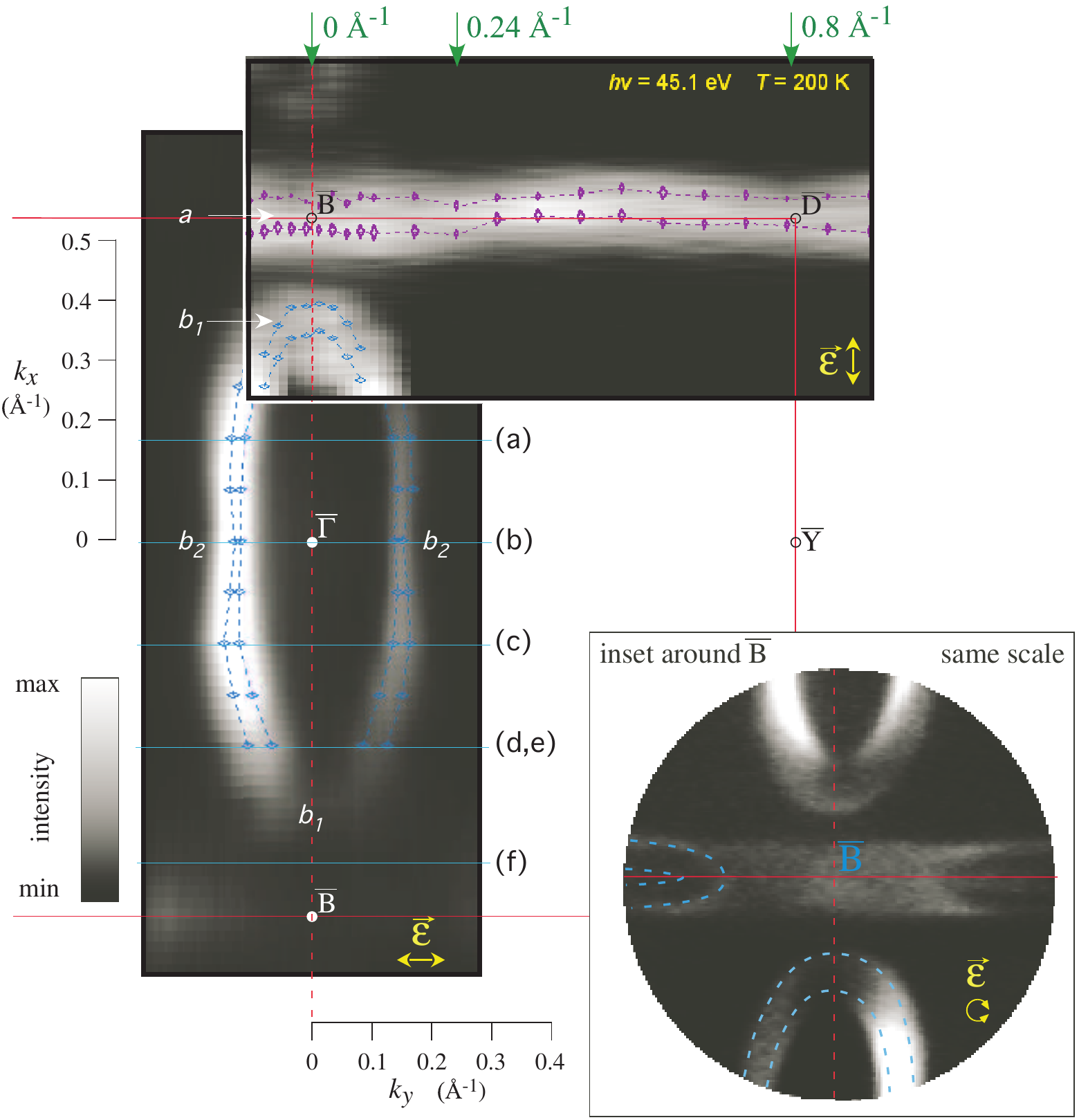}}
\caption{Fermi surface map of ZrTe$_3$ as measured by ARPES at $h\nu=45.1$~eV and $T=200$~K. The global scale is given and the crystal symmetry is indicated by the 2D Brillouin zone boundaries. The features marked {\em "a", "$b_1$"} and {\em "$b_2$"} correspond to Figs.~\ref{disp_GB} and~\ref{disp_GY}. The two panels have been measured with linearly polarized light in different directions as indicated by arrows. Markers show the position of MDC peak fits. Horizontal lines (a-f) correspond to cuts shown in Fig.~\ref{3DFS} and arrows at the top to Fig.~\ref{CDW_gap}. The inset shows a higher resolution data set around $\Bbar$ acquired using elliptically polarized light.}
\label{FSM}
\end{figure}

An orthogonal map through $\Gbar$ in the direction of  $\Ybar$ is shown in Fig.~\ref{disp_GY}. Feature {\em "c"} is again visible as a double peak (EDC in the right section of the figure). The most prominent feature {\em "$b_2$"} is a steeply dispersing band with a Fermi crossing at $k_y = 0.15\ \invA$. Due to the strong intensity differences of the two parallel dispersing copies, its splitting is revealed only by careful peak fitting as shown on the top part of the figure for an MDC at $E_F$ and by tracing the same band along the Fermi surface. 

The Fermi surface map (Fig.~\ref{FSM}) was measured by acquiring dispersion panels at various tilt angles away from the high-symmetry plane. The intensity at $E_F$ was extracted by integrating over a window of 40~meV around $E_F$. The two panels of Fig.~\ref{FSM} were thus measured with orthogonal photon polarizations as indicated.  The data were not symmetrized or distorted. At $T=200$~K, the electronic structure is still unaffected by the CDW transition~\cite{yokoya05,perucchi04} and the Fermi surface is representative of the unmodulated high-temperature unit cell.

\begin{figure}[thb!]
\centerline{\includegraphics[width = 0.38\textwidth]{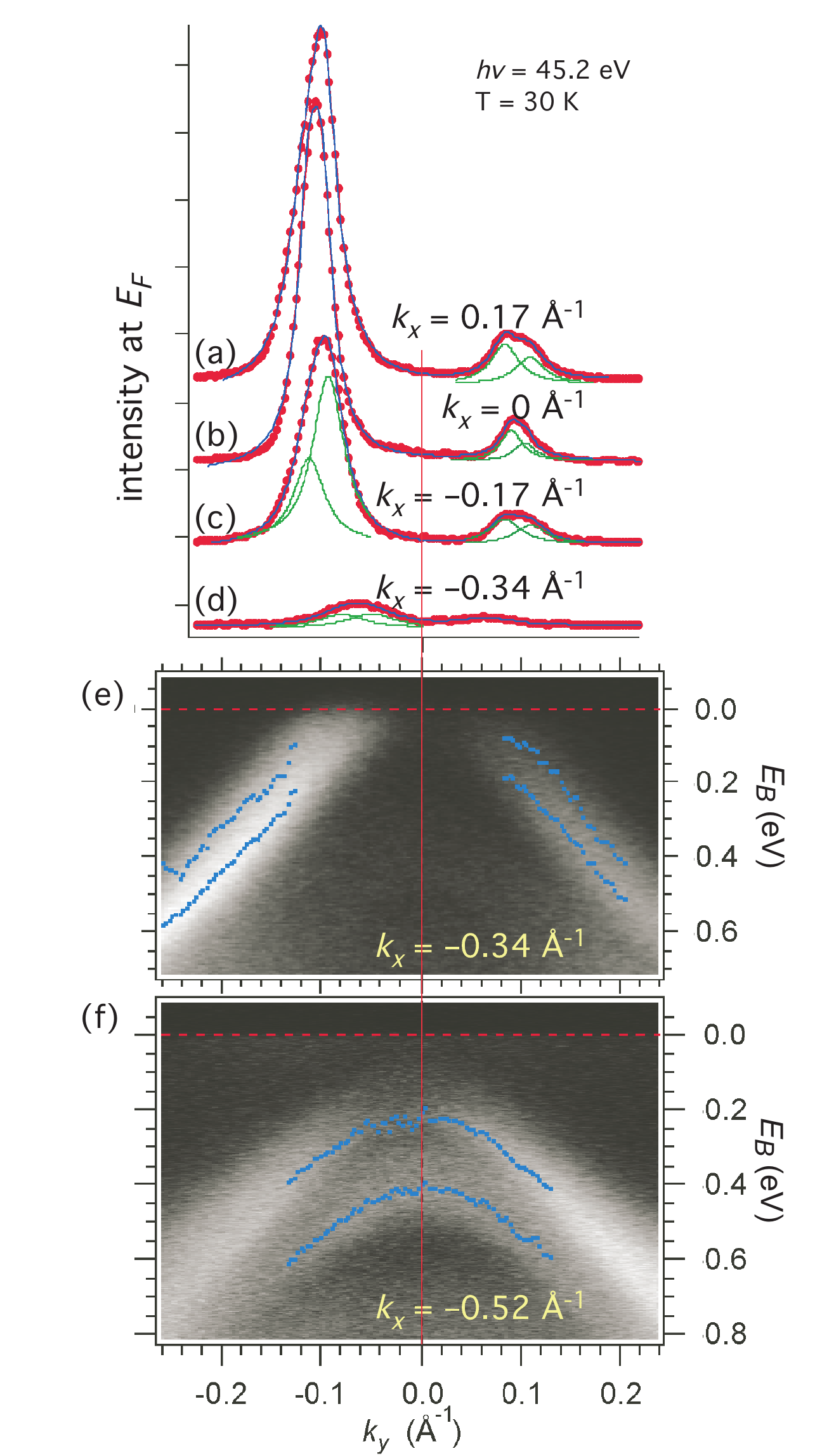}}
\caption{(a-d) Momentum distribution curves at $E_F$ for cuts parallel to $k_y$ at various $k_x$ as indicated by horizontal lines in Fig.~\ref{FSM}. Solid lines are the resulting curves from peak fitting. (e) and (f) ARPES dispersion maps parallel to $k_y$ at $k_x$ as indicated. The peak positions from peak fitting are marked by dots. Map (e) coincides with curve (d).}
\label{3DFS}
\end{figure}

The locus of features {\em "a", "$b_1$"} and {\em "$b_2$"} is indicated in the Fermi surface map in Fig.~\ref{FSM}. The identification of these features enables us to attribute the small pocket {\em "a"} to the small electron pocket of the q1D Fermi surface. The intensity is seen as a wide band on both sides of the BZB. Features {\em "$b_1$"} and {\em "$b_2$"} form the more corrugated 3D Fermi surface. 

The pocket of intensity of the q1D sheet (feature {\em "a"}) is visible all along the line $\Bbar$ - $\Dbar$. The centre between the two MDC peak positions forms a slightly curved line although it should follow the straight line of the BZB. We assign this artificial curvature to slight motions of the beam spot as the sample is rotated. The observation of a dispersion (see also Fig.~\ref{CDW_gap} below) allows the determination of the two peak positions of the Fermi wave vectors from the MDCs. The inset displaying high resolution data around $\Bbar$ reveals that at least one Fermi surface sheet in the q1D pocket does not reach all the way to the $\Bbar$ point as a closing arc is visible. However, a high intensity is seen in a region everywhere along the BZB including $\Bbar$. Inside the sharp Fermi arc, towards the edges of the plot inset, a further Fermi surface sheet is observed which yields intensity in a sharp line following the BZB.

The 3D sheet ({\em "$b_1$"} and {\em "$b_2$"}) has the shape of a flattened oval. The splitting leads to two concentric copies. The Fermi wave vectors have been determined by peak fitting  four Lorentzian peaks to the MDCs at $E_F$ and the resulting peak positions are indicated by markers in Fig~\ref{FSM}. The momentum splitting at location {\em "$b_2$"} is reduced to almost zero. This is due to the steep dispersion of the band as seen in Fig.~\ref{disp_GY}. 
The splitting becomes readily visible in its evolution along the Fermi surface as shown in Fig.~\ref{3DFS}. The MDCs shown in the top sections (a-d) are well described by peak fitting four Lorentzian peaks of equal width. In panels (e) and (f) showing maps along $k_y$ for values of $k_x$ where the FS sheet closes, peak fitting was performed on the EDCs, identifying two parallel copies of the band and allowing a splitting of $120-180$~meV to be determined. Note that panel (f) corresponds to the occupied part of feature {\em "$b_1$"} in Fig.~\ref{disp_GB}. The intensities vary due to the different directions of the polarization of the light. In particular the contrast to another band obscuring the view in Fig.~\ref{disp_GB} is greatly enhanced.

\section{Origin of the splitting}
\label{sect_split}

The splitting of both the band at $\Gbar$ {\em "c"} and of the 3D Fermi surface by 100-200 meV is not expected in the bulk electronic structure. Both bands are derived from highly directional Te $p$-orbitals with some hybridization with Zr $d$ states.\cite{felser98} The low-symmetry crystal structure does not permit a degeneracy of these bands (except for the Kramers degeneracy due to the inversion symmetry, which forces spin-up and spin-down to be degenerate). Possible explanations of a splitting of this order of magnitude include ({\em i}) ferromagnetic symmetry breaking leading to an exchange splitting or ({\em ii}) spin-orbit (SO) splitting of the kind found in the closed $p$-shells of atoms. The former ({\em i}) would require the sample to be ferromagnetic, which has not been observed. The latter ({\em ii}) was proposed by  Pacile {et al.} in the related ZrSe$_3$, HfSe$_3$ and ZrS$_3$.\cite{pacile07}  In these semiconducting systems a band at $\Gamma$ is observed in ARPES spectra that disperses like feature {\em "c"} in our data and shows a similar splitting. The arguments in favour of SO splitting are (a) the order of magnitude of the splitting which matches atomic expectation for the chalcogenide $p$-shell and (b) an increase of the splitting in the selenide compared to the sulphide. This splitting would require, however, that the underlying band has a degeneracy which could be lifted by the SO-interaction. Other origins of the splitting could be of a geometric nature, such as ({\em iii}) bi-layer splitting or ({\em iv}) inhomogeneity of the surface leading to a superposition of features from differently cleaved or clean and dirty portions. We exclude this last possibility ({\em iv}) since the splitting and intensity ratios of the features were observed reproducibly on various cleaves in different sample orientations and in different laboratories.

 \begin{figure*}[thb!]
\centerline{\includegraphics[width = 0.9\textwidth]{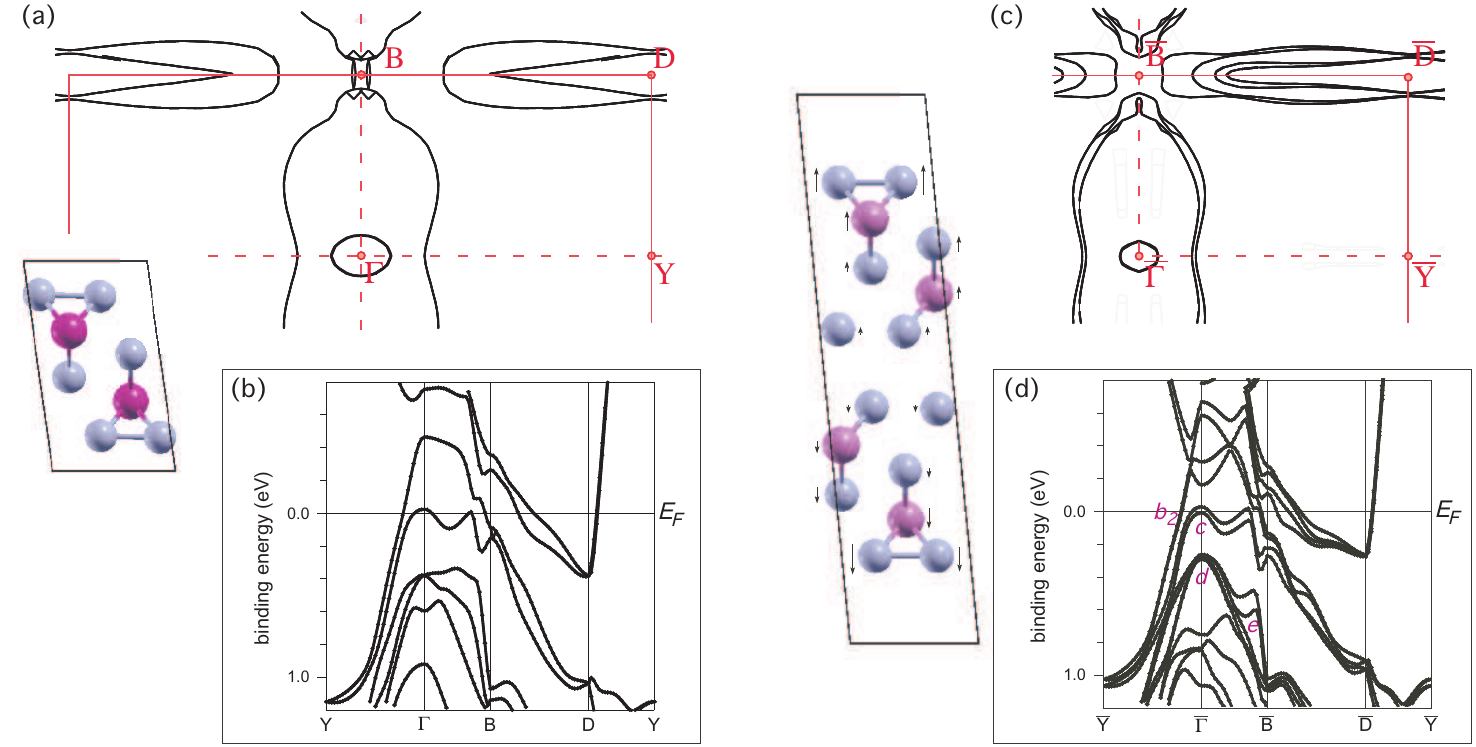}}
\caption{Fermi surface and band structure calculations of (a-b) bulk ZrTe$_3$ and (c-d) a two-layer slab of slightly distorted ZrTe$_3$. The unit cells for the {\em ab-initio} are shown schematically. In the slab unit cell, the displacements with respect to the bulk configuration are indicated by arrows.}
\label{theory}
\end{figure*}

In order to further identify the character of the observed bands, we performed a first-principles band structure calculation. The result of the Fermi surface and band dispersion of bulk ZrTe$_3$ in the $a^*$-$b^*$ plane are shown in Fig.~\ref{theory}(a) and (c). The agreement of this calculation and the experiment is rather poor, although we may assign most of the bands. Feature {\em "d"} at $\Gamma$ and the 3D Fermi surface are readily identified as the band at 0.5~eV binding energy (in agreement with the data) and the large hole pocket extending in the $\Gamma$-B direction, respectively. The q1D Fermi surface is found over a large portion along the BZB from B-D and consists of two sheets due to the two inequivalent atoms which form the Te-Te chains. The band {\em "c"} is assigned to the hole pocket around $\Gamma$, thus incorrectly crossing the Fermi level. Such a small pocket with few carriers could easily be an artifact of the calculation and has little significance for the total energy. The calculations were performed using a scalar-relativistic treatment of the SO interaction, thus any degeneracy-lifting due to the SO energy should be visible in the calculation. Overall, the agreement of this theory with the experiment is so poor that even a precise assignment of all bands is hard to achieve.

An inspection of the calculations shows that the bands in question are non-degenerate (except for the Kramers degeneracy). The splitting due to SO interaction ({\em ii}) must therefore be ruled out. Of the scenarios listed above we are left with ({\em iii}) a bi-layer splitting due to "almost but not quite" equivalent atoms in the effective unit cell. A surface reconstruction which could lead to an effectively larger surface unit cell and hence inequivalent atoms has not, however, been observed by LEED. We propose to consider an out-of-plane relaxation leading to an effective bi-layer situation near the surface. To model the surface we construct a repeated slab of two inequivalent ZrTe$_3$ layers. The slabs are separated by a 10 Bohr thick vacuum spacer and the structure is relaxed in Wien2k. This leads to a widening of the layers and the two layers move slightly apart as shown in the schematic unit cell in Fig.~\ref{theory}(b). The electronic structure in this stable configuration is strictly two-dimensional. The real surface configuration of a cleaved sample is probably close to  the bulk configuration up until the second-outermost layer. The outermost layer, deprived of its neighbours, can relax slightly into the vacuum and expand. The probing depth of ARPES at $E_k=45$~eV is likely to show only features from the outermost or two outermost layers. This surface configuration could in principle be modelled by a large slab calculation comprising many unreconstructed and reconstructed layers. However, such a calculation is beyond the scope of this paper. We believe that we capture the most important features with our simple bi-layer model.

The calculated Fermi surface (FS) and  band structure of the bi-layer are shown in Fig.~\ref{theory}(c) and (d). The 3D FS sheet is split into two concentric sheets. The q1D FS sheets now consist of four sheets, all of which are parallel to $\Bbar$-$\Dbar$ and fulfill good nesting conditions. The small hole pocket around $\Gbar$ still appears in the calculation, but the dispersion reveals that only one of the split bands is slightly unoccupied. Except for this mismatch of binding energy, the overall agreement of this calculation with the experiment is much better. In particular, feature {\em "d"} and the hook-like feature {\em "e"} close to $\Bbar$ are recognized in the data (cf. Fig.~\ref{disp_GB}). The splitting of the 3D Fermi surface is well-reproduced in this calculation. The splitting of feature {\em "$b_1$"} at $\Gbar$ is seen in the calculation, although it is smaller (50 meV) than that observed experimentally (100 meV). In agreement with the experiment, the splitting increases to about twice this value along the dispersion $\Gbar$-$\Bbar$. 

Based on the agreement of the two-layer slab calculation with our experimental data, we consider the surface of cleaved ZrTe$_3$ to be strongly relaxed in the out-of-plane direction. The detailed atomic structure cannot be derived from our spectroscopic data, but the observed splitting suggests a configuration with a reconstructed surface layer which is almost equivalent to the bulk layers. This conclusion leads to the interesting consequence that other photoemission studies on these surfaces (refs.~\onlinecite{yokoya05,starkovicz07,pacile07}) may also have probed a special surface configuration of the material. 







\section{The CDW gap}
\label{sect_CDW}

\begin{figure*}[thb!]
\centerline{\includegraphics[width = \textwidth]{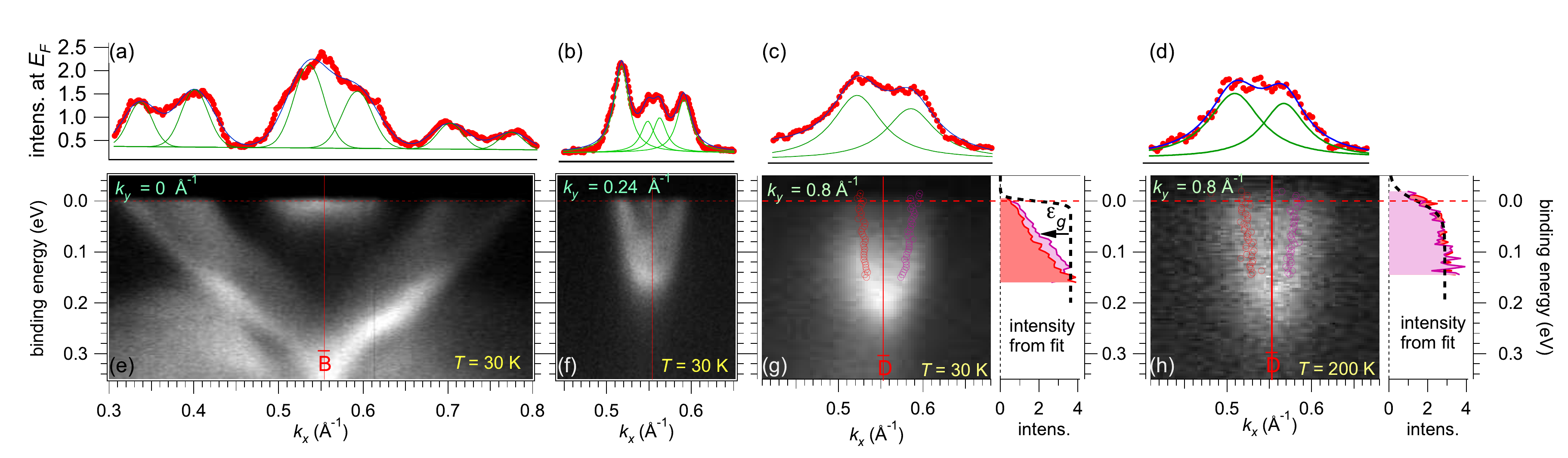}}
\caption{ARPES dispersion maps at three different position along the Brillouin zone boundray BZB from $\Bbar$ to $\Dbar$ at low $T=30$~K (e-g) and $T=200$~K (h). The Fermi wave vector on either side of the BZB is determined by peak fitting as shown in (a-d) of MDCs at $E_F$. In panels (g-h) the dispersion is traced by fits to the MDCs by two Lorentzians of equal width. The intensities of these fits are shown to the right of the panel together with a Fermi occupation function (dashed line) at the corresponding temperature. The half-height of the intensity curve at $T=30$~K (h) corresponds to the magnitude of the gap $\varepsilon_g$ and is marked by an arrow. The data sets (a-b) and (e-f) were acquired with a higher resolution spectrometer and elliptically polarized light, whereas (c-d) and (g-h) were acquired with linearly polarized light.}
\label{CDW_gap}
\end{figure*}

Peierls-type instabilities are released by a momentum dependent gap in the electronic structure at $E_F$ corresponding to the incommensurately modulated structure.
The gap of ZrTe$_3$ has previously been observed,\cite{yokoya05,starkovicz07} but only as a reduction of spectral weight at the Fermi level within the intensity on the BZB. The dispersion of the electron band could not be resolved. The higher resolution of the spectrometers used in the current study allows the band to be seen as a parabolic dispersion along $k_x$ for all sections along $\Bbar$ -$\Dbar$. As the temperature is lowered and spectral weight is shifted away from the Fermi level, the band dispersions remain traceable up to $E_F$, although the intensity is reduced within the gap. Fig.~\ref{CDW_gap} shows three sections along $k_x$ at $\Bbar$ (equivalent to Fig.~\ref{disp_GB}(a)), near $\Bbar$ at $k_y = 0.24\ \invA$ and at $\Dbar$.  At least one electron-like band is observed at $\Bbar$, two at $k_y = 0.24\ \invA$ and again one at $\Dbar$. 

The Fermi surface is thus partially gapped, leaving ample sheets for both metallic conductivity below $T_{CDW}$ and superconductivity. The 3D sheet does not show any change with temperature, the q1D sheet has a fully developed gap in a large region around $\Dbar$. On approaching $\Bbar$, the gap is reduced and no gap is seen for $k_y<0.2\ \invA$. The inset of Fig.~\ref{FSM} shows a closing arc of one sheet of the FS at $k_y =0.16\ \invA$. At $\Bbar$ a high intensity is, however, still present. We assign this intensity to another FS sheet that is not identified by the calculations. This FS sheet at $\Bbar$ deserves special attention as Yokoya {\em et al.}\cite{yokoya05} have found a spectacular increase in spectral weight close to $E_F$ at low temperature in this position. In our data the MDC at $E_F$ of the intensity at $\Bbar$ is compatible with a single Fermi surface crossing (Fig.~\ref{CDW_gap}(a)). The band is not observed when the photon polarization is along $b^*$ (see Fig.~\ref{FSM}), thus it is even with respect to the $b^*$-$c^*$-plane. It was suggested that a van Hove singularity (vHs) could be present at this point at $E_F$ due to the fact that the 3D and q1D FS sheets touch each other.\cite{stowe98,felser98} Any involvement of the 3D FS sheets can be ruled out, however, since its Fermi wave vectors are found to be $k_x = 0.32\ \invA$ and $k_x= 0.4\ \invA$ respectively for the two split sheets, at a safe distance from the $\Bbar$-point ($k_x = 0.55\ \invA$). Our calculations fail to predict a Fermi surface sheet at $\Bbar$ and cannot assign the observed band. The hybridization of the Te $p_x$-band with the band dispersion from feature {\em "c"} could still lead to a small vHs (cf. Fig.~\ref{theory}(d)).

In the region $k_y > 0.2\ \invA$, where the CDW gap was observed in previous studies,\cite{yokoya05,starkovicz07} the band dispersion and intensity distribution is determined by a peak fitting analysis of the MDCs. In a large region around $\Dbar$, the gap is fully developed at $T=30$~K.\cite{yokoya05} Surprisingly, the band dispersion remains unambiguously traceable up to $E_F$ as shown in Fig.~\ref{CDW_gap}(f-g). The shift of spectral weight away from $E_F$ is seen as a linear increase of the band intensity over a binding energy interval of 150~meV. This compares to a constant intensity over the whole binding energy range derived from a similar analysis at room temperature (Fig.~\ref{CDW_gap}(h)). The expected back-folding of the band due to the new periodicity is not observed and the gap value is determined as the binding energy where the band acquires half of its maximum intensity  $\varepsilon_g = 65\pm10$~meV. An upper band apex  (whose binding energy would correspond to the gap value) is not found.

ZrTe$_3$ is a rather special CDW system since the component $q_{CDW}{}^{(a^*)}=0.07$ along $a^*$ is very small and correspondingly the wavelength of the modulation along the chains is very long ($\lambda \simeq 80$~\AA). This property is conserved in the cleaved surface of ZrTe$_3$ and we determine $2k_F=0.06\ \invA=0.11 \cdot a^*$, slightly larger than suggested by the bulk superstructure. This nesting suggests a CDW in the surface layer as $(\overline{q}^x,\overline{q}^y)= (0.11,0)$. The long wavelength and the finite correlation length of the  Peierls modulation, even well below the transition temperature $T_{CDW}=63$~K, lead to a smearing of the band folding in momentum space. Since the Peierls distortion only creates a slight modulation in the underlying crystal structure, it is not surprising that the back-folded band is not visible in the data. However, the removal of spectral weight from the Fermi level is still assigned to a partial CDW gapping of the Fermi surface.



\section{Conclusions}

We have presented a detailed Fermi surface map and band dispersion study of ZrTe$_3$. The basic features of the Fermi surface, i.e. a large 3D body around $\Gbar$ and a q1D sheet along the Brillouin zone boundary $\Bbar$-$\Dbar$ are consistent with previous results, but the 3D Fermi surface as well as a band at $\Gbar$ are found to be split by 100-200 meV. This splitting is explained by a surface relaxation leading to a bi-layer splitting of the electronic structure. 


The q1D FS sheets are seen all along the BZB $\Bbar$-$\Dbar$. The band is traceable by peak fitting with a Fermi wave vector $2k_F = 0.06\ \invA$. This opening of the electron pocket is seen everywhere along the BZB, although at least one sheet of the Fermi surface closes in an arc at $k_y=0.15\ \invA$. A second sheet is identified inside the nested sheet and at $\Dbar$ they are degenerate. Very good nesting conditions are therefore found all along $\Bbar$-$\Dbar$. The CDW gap is identified as a shift of spectral weight away from $E_F$ in a gap of $\varepsilon_g = 65\pm10$~meV for $k_y>0.2\ \invA$. This gap is not seen as a back-folded band, due to the long wavelength and finite correlation length of the Peierls distortion. 

This leads to the conclusion that the surface of ZrTe$_3$, as well as other MX$_3$ systems, does not have a simple bulk termination. Even though the crystal probably cleaves in the van der Waals bonded gap between Te layers and the surface termination has no dangling bonds, the outermost layer relaxes into a configuration which is slightly different from the situation in the bulk. The electronic levels of this surface system split in a fashion similar to a bi-layer splitting. All observations by highly surface sensitive techinques, in particular the VUV ARPES studies performed by Yokoya {\em et al.}, Starkovicz {\em et al.} and Pacile et al.  (Refs. \onlinecite{yokoya05}, \onlinecite{starkovicz07} and \onlinecite{pacile07}), and probably also the studies on NbSe$_3$ by Sch\"afer {\em et al.},\cite{schaefer03} have therefore probed a special surface system of reconstructed MX$_3$ chains, which closely resembles the bulk but is slightly modified. Since a small modification of the atomic arrangement can drastically influence the CDW, this opens up the possibility to deliberately manipulate the CDW material on its surface and thus study the CDW phase transition in new and controlled systems.

\section*{Acknowledgements}
We would like to thank A. Chainani and T. Yokoya for help at the start of experiments. We have gratefully used data analysis routines from A. Ino, M. Miura and F. Baumberger. Inspiring discussions of the data and calculations are acknowledged with M. Grioni, J. Osterwalder, L. Patthey and P. Aebi. We thank W. Welnic for repeating some of the calculations for confirmation and A. Petrovic for careful reading of the manuscript. The experiment was performed at the Hiroshima Synchrotron Radiation Centre (proposal no. 06-A-17) with additional data acquired at the Surface and Interface Spectroscopy beamline at the Swiss Light Source. {\em Ab-initio} calculations were performed using the Wien2k package. One of us (MH) would like to thank the Japanese Society for the Promotion of Science (JSPS) for financial support. Support by the Fonds National Suisse pour la Recherche Scientifique through Div. II and the Swiss National Center of Competence in Research MaNEP is gratefully acknowledged.

\bibliography{ZrTe3}

\end{document}